# Comparison of simulated backgrounds with in-orbit observations for HE, ME and LE onboard *Insight-HXMT*

Juan Zhang[1] • Xiaobo Li[1] • Mingyu Ge[1] •
Haisheng Zhao[1] • Youli Tuo[1,2] • Fei Xie[3] •
Gang Li[1] • Shijie Zheng[1] • Jianyin Nie[1] •
Liming Song[1] • Aimei Zhang[1] • Yanji Yang[1] •
Yong Chen[1]

**Abstract** *Insight-HXMT*, the first X-ray astronomical satellite in China, aims to reveal new sources in the Galaxy and to study fundamental physics of X-ray binaries from 1 keV to 250 keV. It has three collimated telescopes, the High Energy X-ray telescope (HE), the Medium Energy X-ray telescope (ME) and the Low Energy X-ray telescope (LE). Before the launch, in-orbit backgrounds of these three telescopes had been estimated through Geant4 simulation, in order to investigate the instrument performance and the achievement of scientific goals. In this work, these simulated backgrounds are compared with in-orbit observations. Good agreement is shown for all three telescopes. For HE, 1) the deviation of the simulated background rate after two years of operation in space is $\sim 5\%$ from the observation; 2) the total background spectrum and the relative abundance of the $\sim 67$ keV line show long-term increases both in simulations and observations. For ME, 1) the deviation of simulated background rate is within $\sim 15\%$ from the observation, and 2) there are no obvious long-term increase features in the background spectra of simulations and observations. For LE, the background level given by simulations is also consistent with observations. The consistencies of these comparisons validate that the *Insight-HXMT* mass model, i.e. space environment components and models adopted, physics processes selected and detector constructions built, is reasonable. However, the line features at $\sim 7.5$ keV and 8.0 keV, which are obvious in the observed spectra of LE, are not evident in simulations. This might result from uncertainties in the LE constructions.

Juan Zhang

Xiaobo Li

Mingyu Ge

Haisheng Zhao

Youli Tuo

Fei Xie

Gang Li

Shijie Zheng

Jianyin Nie

Liming Song

Aimei Zhang

Yanji Yang

Yong Chen

zhangjuan@ihep.ac.cn

[1]Key Laboratory of Particle Astrophysics, Institute of High Energy Physics, Chinese Academy of Sciences, Beijing 100049, China

[2]University of Chinese Academy of Sciences

[3]INAF-IAPS, via del Fosso del Cavaliere 100, I-00133 Roma, Italy

## 1 Introduction

Due to atmospheric absorption, X-ray radiation of astrophysical sources needs to be detected in space. However, satellite-borne detectors suffer enormous space radiation including cosmic rays, diffuse X-rays, solar flares, the albedo of the Earth, charged particles trapped in the radiation belts and so on. Besides causing damage to sensitive detectors, the space radiation also results in background events during scientific observations of target sources. The space-induced background varies with each instrument, according to the detector type and operation orbit (e.g. Jahoda et al. 2006; Rothschild et al. 1998; Fukazawa et al. 2009; Tawa et al. 2008). The in-orbit background of each space instrument has to be estimated before the launch in order to optimize the instrument design and to investigate how well it can fulfill the scientific goals. Geant4 (Agostinelli & Geant4 Collaboration 2003; Allison et al. 2006, 2016) is a general toolkit to simulate the interaction of particles with matter. It is widely used in nuclear physics and accelerator physics, as well as medical science and space science. Currently, it is also a popular tool to predict the in-orbit backgrounds of space instruments (e.g. Tenzer et al. 2010; Perinati et al. 2012; Fioretti et al. 2016; Weidenspointner, Pia, & Zoglauer 2008; Campana et al. 2013; Xie & Pearce 2018). The accuracy of simulated backgrounds could be examined by in-orbit observations obtained after the launch. This kind of examination can validate the simulation methods and the assumed models. For example, Mizuno et al. (2004) validated the cosmic ray background flux models based on a *GLAST* balloon fight experiment. And Odaka et al. (2018) verified that a simulation process of proton-induced radioactivation background could describe background measurements of Hitomi/Hard X-ray telescopes very well through comparison between simulations and measurements.

As the first Chinese X-ray astronomical satellite, the Hard X-ray Modulation Telescope (*HXMT*), also named *Insight-HXMT*, was launched into a low-Earth orbit with an altitude of ∼550 km and an inclination of ∼ 43° on 15th June 2017 (Li et al. 2018). It aims to scan the Galactic plane for new sources and to study fundamental physics of X-ray binaries (Li 2007; Zhang et al. 2020). The three scientific payloads, the High Energy X-ray telescope (HE, 20–250 keV), the Medium Energy X-ray telescope (ME, 5–30 keV) and the Low Energy X-ray telescope (LE, 1–15 keV), are slat-collimated instruments and co-aligned. Each has different sized field of view (FOV) and rotation angle, with the aim to subtracting in-orbit background using the combined FOV method (Jin et al. 2010). A mass model was built in the framework of Geant4 to estimate the in-orbit backgrounds of these three payloads (Xie et al. 2015; Li et al. 2015). The simulation results showed that the estimated background flux of HE was comparable to the background measurements of *RXTE/HEXTE* (Xie et al. 2015). *RXTE/HEXTE* was selected for comparison because it was also a slat-collimated instrument, with the same kind of scintillators (NaI/CsI) and a similar operation orbit. A big amount of data on *Insight-HXMT* background observations has been accumulated so far. In this work, we analyze the background measurements of HE, ME and LE, and make comparisons with previously simulated results, with the motivation to examine the mass model. It is worth noting that the backgrounds given by simulations are an average estimate after a long-term operation in space and it cannot be used in data analysis for any practical observations.

The paper is structured as follows. In Section 2, the mass models of the *Insight-HXMT* HE, ME and LE are briefly introduced. This is followed by Section 3, where background observations are analyzed and results between simulations and observations are compared. The discussion and conclusion are given in Section 4.

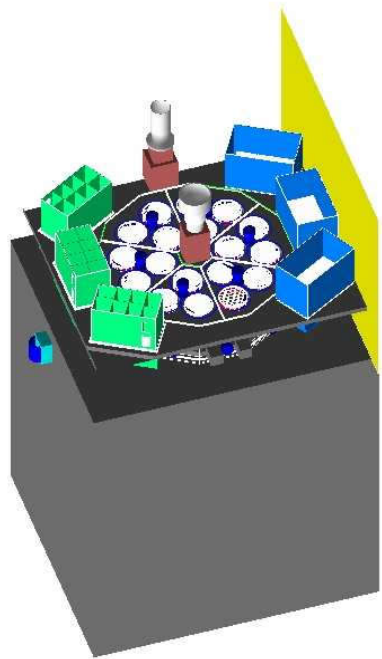

**Fig. 1** The whole structure of Insight-HXMT. HE is surrounded by ACDs. ME consists of three boxes which are in blue color. LE consists of three boxes in green. The gray cubic at the bottom is the service module.

## 2 Mass Modeling

The mass model of *Insight-HXMT* has been built under the framework of Geant4 Version 9.4.p04 (Xie et al. 2015) and is updated to Version 10.05.p01 currently. Three objects that needed to be specified are detector constructions, primary particles and physics processes, and they are described as follows.

2.1 Detector Constructions

The whole satellite and detailed structures of HE, ME and LE are shown in Fig. 1, which is generated by Geant4 visualization. HE is in the middle of the platform. It is surrounded by 18 anti-coincident detectors (ACDs). ME and LE are at the two edges of the platform. Both ME and LE have three boxes, which are illustrated in blue and green colors respectively.

HE is equipped with NaI(3.5 mm)/CsI(4 cm) scintillators. The collimators of HE are made of aluminum alloy and tantalum. ACDs above and around HE are used to veto charged particles in space. With pulse shape discriminator (PSD), signals in the CsI could be distinguished from those in the NaI. This can be used to reject high energy gamma rays and charged particle backgrounds. HE consists of 18 detector modules, one of which has a blind FOV collimator, two have large FOV ($5.7° \times 5.7°$) collimators, and the other 15 have small FOV ($1.1° \times 5.7°$) collimators. Arrangements of these collimators are shown in Fig. 2. The cut-away view of one small FOV detector module is presented in Fig. 3. These figures are generated by Geant4 and illustrate how the mass model of HE is built in detail.

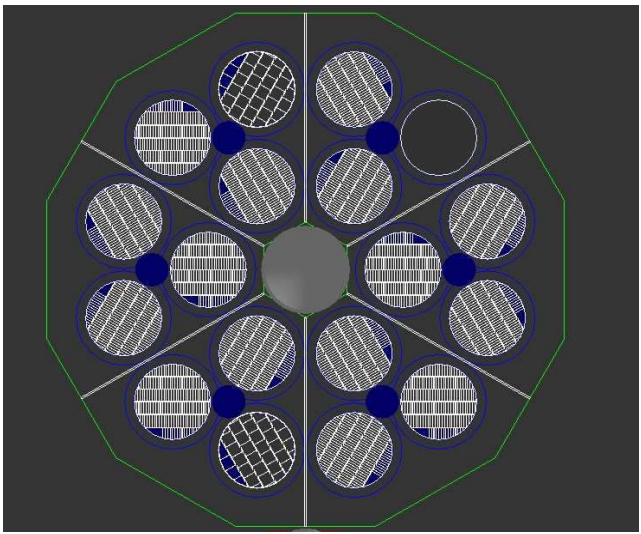

**Fig. 2** Top-down view of the collimators of 18 HE modules. The white grids are the collimator slats. The top-right one (a white empty circle) is the blind FOV collimator. The two with sparse grids are large FOV collimators. The other 15 are small FOV collimators. This figure is generated by Geant4 visualization directly.

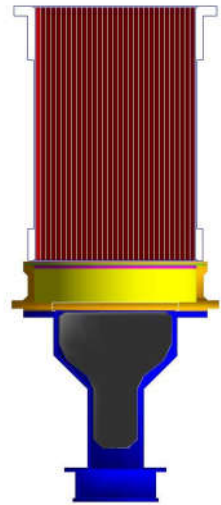

**Fig. 3** The mass model of one HE small FOV detector module. The yellow layer in the middle is 4 cm CsI. The layer above is 3.5 mm NaI, which is in magenta. The green layer above NaI is the 1.5 mm beryllium window. All of these are installed inside the crystal holder, which is aluminum alloy and is shown in orange color. Above the crystal holder are collimated slats, which are installed inside an aluminum alloy cylinder. The signals in NaI/CsI crystal are read out by PMT, which is illustrated below the crystal holder.

Both ME and LE are silicon detectors (Zhang et al. 2020). Three boxes of ME are placed with a rotation angle of $120°$ relative to each other. The sensitive detectors, which aim to detect X-rays from 5 keV to 15 keV, are Si-PINs with thickness of 1 mm and covered by 50 $\mu$m beryllium window. Each box of ME has 576 detector pixels. A group of 32 pixels is read out through one ASIC. There are 18 ASICs in each ME box. Silver glue with a thickness of $\sim$14 $\mu$m is used to fix the Si-PIN pixels on the bottom ceramics. The ME collimators are made of aluminum alloy and tantalum. There are three different FOVs in each box. The large FOV has a size of $4° \times 4°$, the small FOV is $1° \times 4°$, and the covered FOV is sheltered by 0.6 mm tantalum. One of the ME boxes is shown in Fig. 4. All the pixels in one box are divided into 18 modules according to the collimators and readout ASICs. Each vertical stripe in Fig. 4 corresponds to one module. In summary there are 15 small FOV modules, 2 large FOV modules, and 1 blind FOV module in each box of ME.

LE uses the swept charge device (SCD) CCD236 (Zhao et al. 2019). It has 96 SCDs, with a sensitive thickness of 50 $\mu$m. LE aims to detect X-rays from 1 keV to 15 keV. There are eight collimators in each box. Every 4 SCDs share one collimator. The grids of collimators are made of aluminum alloy, Al7075, which contains about 90.3% pure aluminum and 5.5% zinc. A layer of 0.2 mm tantalum is pasted around the collimator, to reduce cosmic gamma-ray background and charged particles. On the top of each collimator, to reduce stray lights, there is a very thin layer of shading film ($C_{22}H_{10}O_4N_2$), with 100 nm aluminum below and above. Between the shading film and the collimator, there is a layer of nickel mesh to fix the film. The collimators of LE have three different FOVs. The large FOV is $4° \times 6°$, the small FOV is $1.6° \times 6°$, and the covered FOV has 1 mm aluminum alloy Al7075 on top. Besides these three different FOVs, there is also a very short collimator above 4 SCDs inside each LE box, which extends the FOV as large as $50 \sim 60° \times 2 \sim 6°$. Given the extended FOV, these 12 SCDs in the three boxes are used as sky monitors. In total, there are 20 small FOV detectors, 6 large FOV detectors, 2 covered FOV detectors and 4 wide FOV detectors in each box of LE.

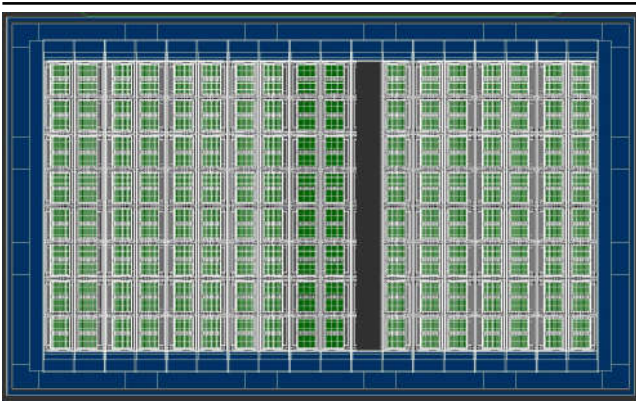

**Fig. 4** The mass model of one ME box, generated by Geant4 visualization. The blocks in green are Si-PIN detectors. Each green block shown in this figure contains two minimum Si-PIN pixels. The white grids above these pixels are the collimator grids. The strip in gray is the shelter that covers the FOV of the blind module. The two strips on the left of this shelter are large FOV modules. The remaining 15 strips are small FOV modules.

All these collimators and detectors are surrounded by an aluminum box, which can shield X-ray and cosmic ray radiation. The whole structure of LE is shown in Fig. 5.

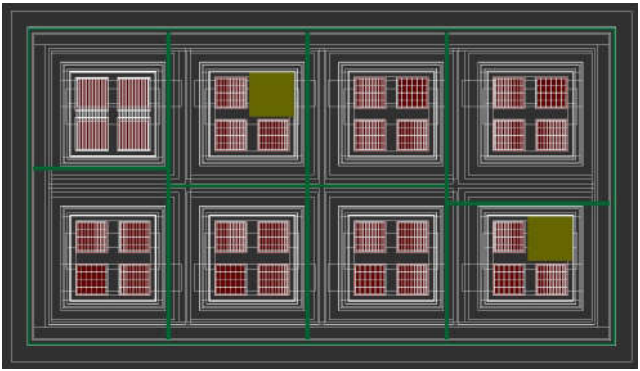

**Fig. 5** The mass model of one LE box, generated by Geant4 visualization. The pixels in red are the SCDs. The grids in white are the collimator grids. The two green blocks are the shelters used to cover the FOV of blind detectors. The top left unit collimator is used as sky monitor.

The service module is located at the bottom of the satellite and is used to hold the fuel and other equipment like attitude control assemblies, etc. It is made of honeycomb boards of aluminum alloy. During the simulations, an increase of the cosmic-ray proton induced backgrounds on HE by dozens of counts is noted if this module is in the whole satellite construction. The raise is mainly from the secondary events generated by the energetic cosmic protons interacted with the massive module. Therefore, the service module is necessary and important in the mass model, as shown to be a hollow gray cubic in Fig. 1. It is constructed according to the real size of its outside dimension with the inner simplified into box structures, and is assigned by the average density derived from the true mass of this module.

2.2 Space Environment

For space environment, various components were included in the simulation, including cosmic X-ray background (CXB), cosmic ray protons and electrons, albedo gamma rays&neutrons, and the trapped protons in South Atlantic Anomaly (SAA) (For more details of each component, please refer to Xie et al. 2015; Li et al. 2008, and references therein). According to the previous simulation, CXB, cosmic ray protons and SAA protons have dominating contributions to the in-orbit backgrounds of *Insight-HXMT*. These components are presented in more details here. An attitude of zenith-pointing is assumed, therefore the incident CXB and primary cosmic rays are obscured by the Earth.

*2.2.1 CXB*

CXB is thought to result from unresolved sources outside the Galaxy. The broken power-law spectrum (Gehrels 1992) is adopted in our simulation,

$$F = \begin{cases} 0.54 \times E^{-1.4}, & E < 0.02\,\mathrm{MeV} \\ 0.0117 \times E^{-2.38}, & 0.02\,\mathrm{MeV} < E < 0.1\,\mathrm{MeV} \\ 0.014 \times E^{-2.3}, & E > 0.1\,\mathrm{MeV} \end{cases} \quad (1)$$

where $E$ is in MeV and $F$ is photons cm$^{-2}$ s$^{-1}$ MeV$^{-1}$ sr$^{-1}$. The $\sim$ 7% normalization fluctuation (Revnivtsev et al. 2003) on angular scales of $\sim$ 1 square deg is ignored. Its position distribution is assumed to be uniform and the direction distribution is isotropic (Dean et al. 2003) except that incident particles go through the Earth are blocked.

*2.2.2 Cosmic Ray Protons*

In a low-Earth orbit, cosmic ray protons outside the SAA region consist of the primary cosmic ray protons, which are modulated by the solar activity and geomagnetic field, and the secondary protons below the geomagnetic cutoff rigidity (Gehrels 1992). To obtain a conservative background estimation, a spectrum corresponding to the minimum solar modulation (Mizuno et al. 2004) and a high geomagnetic latitude is adopted, as shown in Fig. 6. The orbital modulation of geomagnetic fields and the east-west effect are ignored. Similar to CXB, an isotropic distribution except for the Earth direction is used.

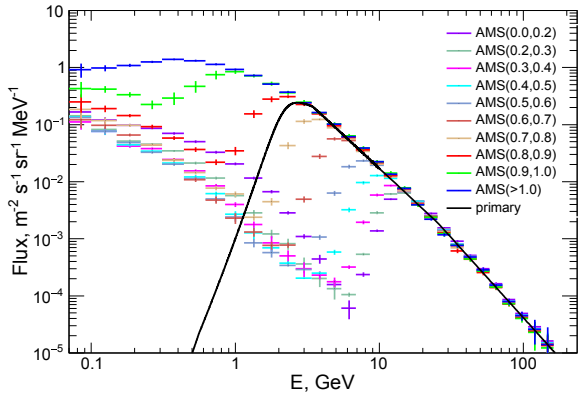

**Fig. 6** The primary cosmic ray proton spectrum used for the *Insight-HXMT* orbit. Also shown are the unfolded downward spectra reported by AMS-01 for different geomagnetic latitudes (Gehrels 1992).

### 2.2.3 SAA Protons

Due to the low-Earth orbit with an altitude of 550 km and an inclination of 43°, *Insight-HXMT* spends $\sim 10\%$ of its operation time in SAA region. SAA induced background is the most dominating component of HE (Xie et al. 2015). The incident SAA proton spectrum is obtained from SPENVIS[1]. The orbit-averaged differential spectrum generated from the AP-8 model for the solar minimum is used, as shown in Fig. 7. The spectrum for solar maximum is also presented for comparison. The direction distribution of SAA protons is assumed to be isotropic and is not blocked by the Earth due to their local origin.

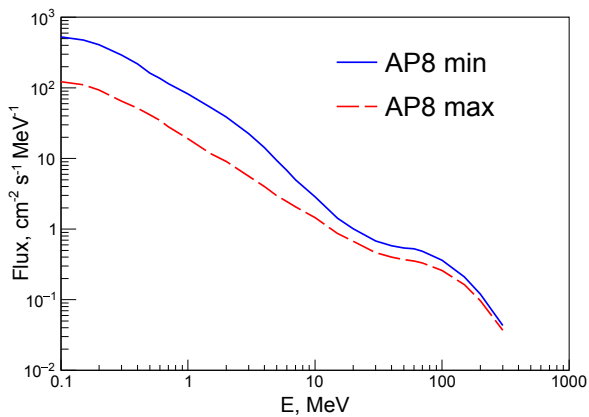

**Fig. 7** The differential spectra of SAA protons for *Insight-HXMT* orbit.

### 2.3 Physics Processes

The element of iodine in HE scintillators could be activated by cosmic ray protons and SAA-trapped protons. Therefore radioactive decay process has to be included in the physics list. Given the energy range of interest of *Insight-HXMT*, 1–250 keV, low energy electromagnetic physics is preferred. Under Geant4 Version 9.4.p04, the Shielding Physics List was chosen in our mass model. As mentioned in Xie et al. (2015), we added radioactive decay process and replaced the standard electromagnetic physics with low energy electromagnetic physics. In Geant4 Version 10.5.p01, the radioactive decay process is already included in the Shielding Physics List, so we only make a minor modification by replacing the standard electromagnetic physics with the livermore electromagnetic physics.

### 2.4 Simulation Output

Sensitive detectors are defined for NaI/CsI of HE, Si-PIN of ME and SCD of LE in the *Insight-HXMT* mass model. The information on these sensitive detectors, such as the deposited energy and its response time, is recorded. These signals are classified into prompt backgrounds and delayed (or radioactive) backgrounds by their response time relative to the incidence. Prompt backgrounds trigger the sensitive detector immediately after the incidence. Radioactive backgrounds could trigger signals on sensitive detector hours, days or even months and years later after the incidence. For these radioactive backgrounds, the deposited energy and the corresponding response time are firstly recorded in Geant4 simulation, then integrated along the radiation history to obtain the final background level.

## 3 Simulation and Observation

### 3.1 Background Observation

Before the launch, some regions in the sky without visible celestial sources had been selected for background observations, due to the lack of the imaging capability. We referred to the INTEGRAL Reference catalog[2] and ROSAT All-Sky Survey Bright Source catalog (Voges et al. 1999) to calculate the significance of each source on HE, ME and LE. The significance of celestial sources is calculated by using signal to noise ratio (SNR), $S/\sqrt{B}$, where $S$ denotes the source counts and

---

[1] https://www.spenvis.oma.be

[2] http://isdc.unige.ch/integral/catalog/39/catalog.html

$B$ the background counts. The simulated background rates are used in this SNR calculation and an exposure time of $10^6$ seconds is assumed. After searching around the medium galactic latitudes, twenty-one directions were chosen, where no sources exceed a significance of 5 within the FOVs of the three instruments. These regions are illustrated in Fig. 8. We define these regions as blank sky directions of *Insight-HXMT*. Positions of these blank sky directions are given in Table 1.

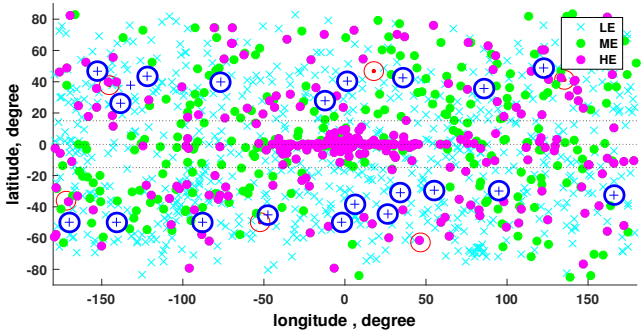

**Fig. 8** Blank sky directions of *Insight-HXMT*, used for background observations. The cross points are the sources visible on LE. The green dots are sources visible for ME, and the magenta dots are sources visible for HE. The 21 blue circles with a cross in the center are the blank sky directions we choose for *Insight-HXMT*. And the background observation directions used for *RXTE/PCA* (Jahoda et al. 2006) are also shown as the red open circles.

Besides these 21 blank sky directions, pointing observations of some certain sources could be also considered as background observations, such as Cas A and PSR B0540–69 for HE. The continuum spectrum of Cas A from 3 to 500 keV could be fitted with a thermal bremsstrahlung plus a power-law components with an index of $3.13 \pm 0.03$ (Wang & Li 2016). Based on this spectrum, the expected contribution of Cas A to HE at 20–300 keV is ∼3 counts per second. For PSR B0540–69, the spectrum of the pulsar plus its wind nebula observed by NuSTAR could be fitted with an absorbed power law with an index of $1.99 \pm 0.01$ (Ge et al. 2019). A rate of ∼2 counts per second is obtained by extrapolating this spectrum to the energy band of HE. Compared with the HE background rate, which is several hundred counts per second (Xie et al. 2015), the contributions of these two sources are negligible. Thus directions of PSR B0540–69 and Cas A could be assumed as blank sky regions to HE.

The first three months after the launch are in performance verification phase. After that, there are routine observations of these blank sky regions. The background observations used in this work are listed in Table 2. To investigate the rapid increase of HE back-grounds in the beginning of space operation, observations of PSR B0540–69 within the first three months are also included. For ME and LE, we averaged several months of data to obtain the orbit-averaged backgrounds, since those instruments have lower background level and shorter effective exposure time. The data analysis and processing of these background observations follows the standard procedure of *Insight-HXMT* software[3] (Zhao et al. 2016). The recommended selection criteria are used, which are also listed in Table 2.

**Table 1** *Insight-HXMT* blank sky directions.

| blank sky # | R.A. (HH:mm:ss.sss) | Dec (dd:mm:ss.sss) |
|---|---|---|
| 1 | 03:18:08.440 | -07:18:53.863 |
| 2 | 03:49:34.839 | -24:22:41.152 |
| 3 | 03:21:14.627 | -57:11:45.715 |
| 4 | 21:48:28.565 | -42:28:20.848 |
| 5 | 23:31:02.978 | -70:19:18.181 |
| 6 | 21:30:22.052 | -23:05:42.197 |
| 7 | 20:48:30.806 | -36:34:32.145 |
| 8 | 03:20:43.597 | +17:33:04.392 |
| 9 | 20:45:16.709 | -12:37:36.317 |
| 10 | 23:02:32.475 | +27:11:58.365 |
| 11 | 21:16:12.821 | +03:49:40.367 |
| 12 | 08:39:48.607 | +04:43:24.787 |
| 13 | 15:37:35.211 | -20:11:04.542 |
| 14 | 17:13:51.057 | +57:28:19.228 |
| 15 | 09:29:54.807 | +05:43:11.503 |
| 16 | 11:46:57.207 | -20:41:17.228 |
| 17 | 15:33:58.625 | -03:26:33.651 |
| 18 | 16:17:22.857 | +20:02:00.034 |
| 19 | 10:06:59.760 | +02:08:11.140 |
| 20 | 09:40:19.010 | +23:05:26.397 |
| 21 | 12:53:06.186 | +68:25:39.179 |

### 3.2 HE Backgrounds

Fig. 9 plots the simulated background spectrum of HE and its constituent components after one year operation in orbit. It is clearly seen that the SAA induced background is the most prominent component. The peak structures of this component at ∼ 31 keV, 56 keV, 67 keV and 191 keV, result from the radioactive decay of iodine isotopes caused by SAA trapped protons. The same structures are presented in the delay background caused by cosmic ray protons. For prompt components induced by cosmic ray protons, CXB, and albedo gamma rays, the fluorescence lines around 57 keV, which originates from HE tantalum collimators, are clearly seen. All of these line structures in background spectrum could be used in the in-orbit calibration of HE (Li et al. 2020).

---

[3] http://hxmt.org/index.php/usersp/dataan

**Table 2** Blank sky observations used in this work.

| | sky regions[a] | observation period[b] YYYYMMDD | label[c] | exposure[d] (seconds) | selection criteria |
|---|---|---|---|---|---|
| HE | PSR B0540–69 | 20170719-20170722 | 20170719 | 45605 | |
| | PSR B0540–69 | 20170913-20170918 | 20170913 | 45748 | |
| | 1-3,10,12,14,15,19,21 | 20171204-20171231 | 20171216 | 59063 | For HE, ME and LE, ELV$> 10°$ |
| | PSR B0540–69 | 20180902-20180914 | 20180902 | 83506 | COR$> 8$ GV |
| | Cas A | 20190713-20190715 | 20190714 | 65464 | SAA_FLAG==0 |
| | Cas A | 20190814-20190816 | 20190815 | 42146 | T_SAA$> 300$ s |
| ME | 1-4,6,8,10-12,14,15,19-21 | 20171102-20171231 | 20171103 | 103320 | TN_SAA$> 300$ s |
| | 2-6,11,14 | 20180611-20181005 | 20180621 | 48658 | ANG_DIST$<= 0.04°$; |
| | 3-6,10,11,14-16,19,20 | 20190429-20190626 | 20190625 | 53286 | for LE, plus |
| LE | 1-4,6,8,10-12,14,15,19-21 | 20171102-20171231 | 20171103 | 67188 | DYE_ELV$> 30°$ |
| | 2-6,11,14 | 20180611-20181005 | 20180621 | 32156 | |
| | 3-6,10,11,14-16,19,20 | 20190429-20190626 | 20190625 | 29633 | |

[a]Numbers in this column indicate the corresponding blank sky directions in Table 1.
[b]The start and end dates of each period. Background observations during these periods are chosen.
[c]The labels in this column are used in Figs. 10, 11 and 12 to indicate that the corresponding background spectra are obtained from the corresponding observation data here.
[d]For the same observation data, HE, ME and LE have different effective exposure time. This is because the data processing has some procedures related to the intrinsic instrument performance itself.

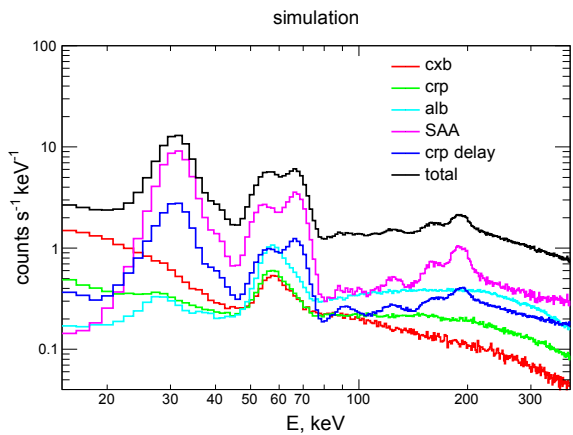

**Fig. 9** Simulated spectrum of the averaged background of HE after 1 year operation in space. This background spectrum consists of prompt background induced by cosmic X-ray background (cxb), cosmic ray protons (crp), the albedo gamma rays of the earth atmosphere (alb), and radioactive background caused by SAA trapped protons (SAA) and cosmic ray protons (crp delay).

After a long term operation in space, the background rate of HE increases, especially at the energy of ∼67 keV, as presented in Fig. 10. Note that the first peak at ∼25 keV in the observation spectra is not a real peak. It is caused by the electronic noise and threshold cutoff (Zhao et al. 2020). These effects are not included in the simulation process, so this peak is not shown in the simulated spectra. Six post-launch periods were chosen to compare the simulation and the observation of long-term variation. For observation, spectra are shown at the periods corresponding to about 1 month (20170719), 3 months (20170913), 6 months (20171216), 1 year (20180902) and 2 years (20190714&20190815) after the launch. Observation data used in this figure are listed in Table 2. For comparison, the simulated HE background spectra after an operation time of 1, 3, 6 months and 1, 2, 4 years are given. The relative increase at ∼67 keV is obvious, which is due to the long decay periods of some iodine isotopes. The relative abundance of ∼67 keV increases quickly especially in the first year of operation, and gradually approaches stable.

The total observed and simulated background rates from 30 to 300 keV on the seventeen large and small FOV modules of HE are shown in Table 3. The average background rate after two year operation is $(538.6 \pm 0.1)$ counts s$^{-1}$ from the observation, and $(563.2 \pm 3.6)$ counts s$^{-1}$ from the simulation. The observed background levels correspond to $3.0 \sim 4.1 \times 10^{-4}$ counts s$^{-1}$ keV$^{-1}$ cm$^{-2}$, and the simulated ones $3.6 \sim 4.3 \times 10^{-4}$ counts s$^{-1}$ keV$^{-1}$ cm$^{-2}$. For the simulated backgrounds of HE, deviations from observations are as large as 20 ∼ 30% in the first three months after the launch. This deviation reduces to ∼16% in the sixth month, ∼10% after 1 year operation and ∼ 5% after 2 year operation. The large deviations in the first three months could be attributed to the fact that *Insight-HXMT* was still in the performance verification phase. On the other hand, as discussed in Xie et al. (2015), the input cosmic proton spectrum we used in simulation corresponded to a high geomagnetic latitude region, which will cause more prompt and delayed backgrounds than that of low geomagnetic latitude regions. While considering that the dominated background component of HE is induced by SAA protons, the deviation finally approaches into a small value after a long term operation in space. In addition, the simulated SAA background contributions are obtained by folding delayed background events output by Geant4 with the averaged SAA irradiation history for *Insight-HXMT* orbit, not the reality history. This will cause some difference in the beginning, but a stable value is approached after a long time accumulation.

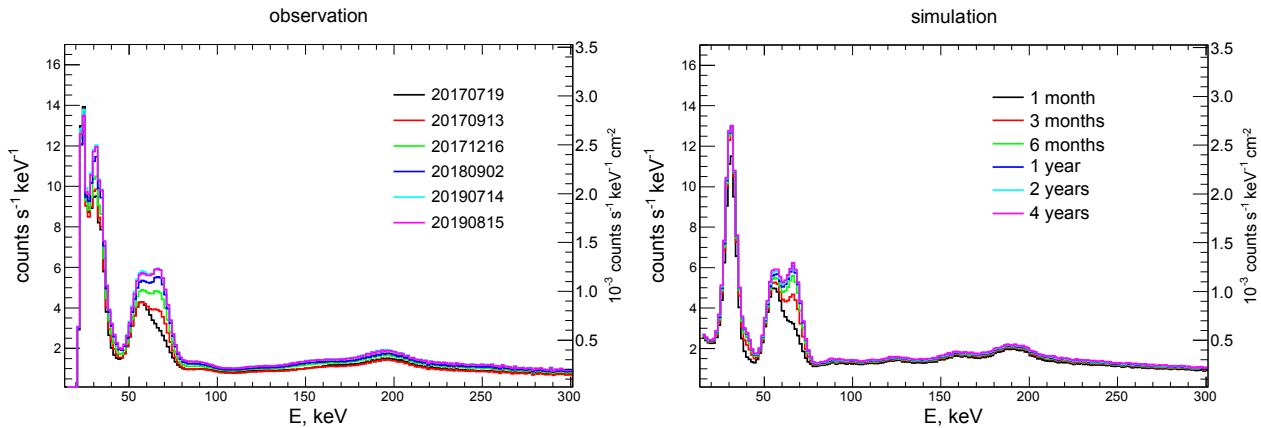

**Fig. 10** The background spectra of HE at different operation periods. Left: the observed spectra from July 19, 2017 to August 15, 2019. Right: the simulated spectra at about 1 month ∼ 4 years later after the launch. For comparison, the spectra shown here include all of the contributions of HE modules except for the covered detector module.

**Table 3** The observed and simulated background rates of HE from 30 to 300 keV.

| period | observation (counts/s) | simulation (counts/s) | deviation (%) |
|---|---|---|---|
| 1 month | 395.2 | 469.2 | 18.7 |
| 3 months | 399.9 | 512.2 | 28.1 |
| 6 months | 461.3 | 536.0 | 16.2 |
| ∼ 1 year | 499.6 | 553.0 | 10.7 |
| ∼ 2 year | 538.6 | 563.2 | 4.7 |

### 3.3 ME Backgrounds

Due to electronic noises presented at low energies, the effective energy range of ME shifts into a bit higher energy band, ∼10-40 keV. The observed backgrounds are shown in Fig. 11, where the three spectra correspond to the operation time of ∼ 6 months (20171103), 1 year (20180621) and 2 years (20190625). The peak around 22.5 keV comes from the fluorescent emission lines of silver, which is an ingredient of the glue between Si-PIN and the ceramics. From these observations, no obvious features present on the long-time variation of ME background spectra. This character could also be inferred from simulation results, which are shown in the right panel of Fig. 11, where the simulated background spectra of ME and its components after 1 year operation are presented. Compared to HE, ME background is not dominated by the delay components. The SAA component contributes less than 10% at the whole energy band, which well explains the long-term stability of the total background.

Given that the geometry size of Si-PIN is quite small, ∼ 12.5 mm × 4.5 mm, and that there are 576 pixels in one box of ME, a method called the "veto between Si-PIN pixels" is used to reduce the background of charged and/or high energy particles in the analysis of simulation data. In this method, events that trigger more than one pixel at a time are discarded, as they are more likely to be caused by cosmic ray protons or high energy gamma rays. This method is tested, as plotted in the right panel of Fig. 11, where thinner lines are backgrounds before veto, and thicker lines correspond to backgrounds after veto. Before veto, the background induced by cosmic ray protons is the dominate component in ME background spectrum, while it is reduced by ∼ 50% after veto between Si-PIN pixels. This veto method hardly has effects on the component of X-rays. The contributions from CXB before and after veto are almost the same. This means the veto method between pixels could reduce the charged particle background efficiently while keeping the same detection efficiency of X-rays.

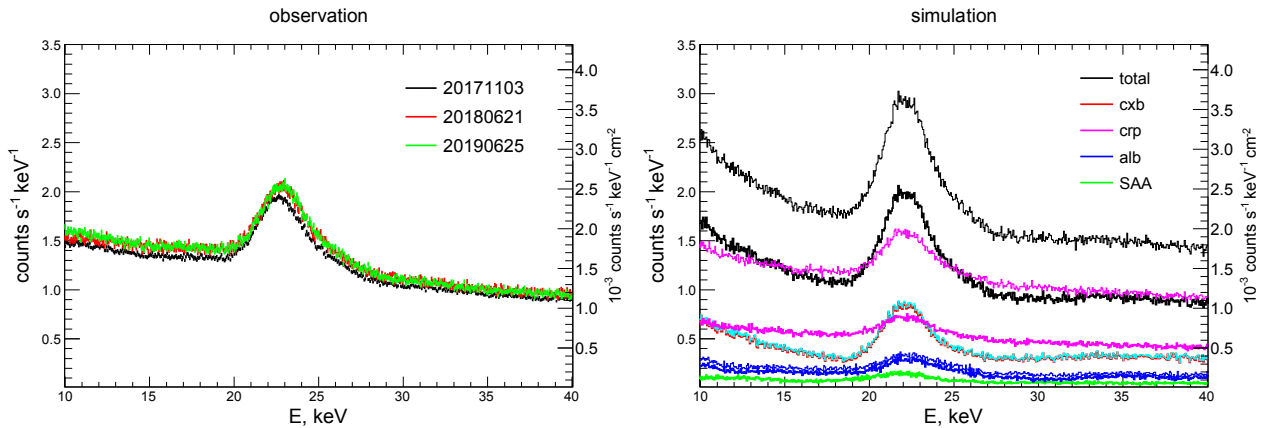

**Fig. 11** Left: The observed background spectra of ME at different operation periods. The black line(20171103) is the spectrum of ∼ 6 months later after the launch, the red line (20180621) 1 year later after the launch and the green line(20190625) 2 years later after the launch. Right: The simulated background spectra of ME. For each component, we plot two background spectra before (the thin line) and after (the thick line) veto. The CXB component before veto is plotted in cyan in order to distinguish it from the line after veto. For the simulation and observation spectra, only small FOV detectors are selected.

The simulated ME background rate after veto from 10 to 40 keV on small FOV detectors is $(35.1 \pm 0.2)$ counts s$^{-1}$ after 1 year operation, which corresponds to a background level of $1.4 \times 10^{-3}$ counts s$^{-1}$ keV$^{-1}$ cm$^{-2}$. The observed rate is $(37.8 \sim 40.2) \pm 0.1$ counts s$^{-1}$, i.e. $1.6 \sim 1.7 \times 10^{-3}$ counts s$^{-1}$ keV$^{-1}$ cm$^{-2}$. The simulated background is about 15% lower than the observed value. In addition, the relative abundance of silver lines given by simulation is slightly different from those given by observations. As for the silver line, the exact thickness of the silver glue layer is unknown and the thickness of this layer below different pixels are not the same. We used an uniform thickness of 14 μm in the simulation. For the background level, note that we do not consider any electronic and readout processes in simulation and that the simulation result is obtained after veto between pixels for each simulated background event. This veto method is an ideal and much more strict selection criterion than the condition of "Grade==0", which is used by default in the observation data analysis procedure. This will cause a lower simulated proton background estimation. In addition, the number of pixels that were selected during observation data analysis and processing could also introduce some difference between simulation and observation as well. Si-PIN pixels with high electronic noise are screened out during observation data analysis. The background counts on the remaining pixels were scaled to all pixels of the 45 small FOV detector modules to obtain the observed rate of ME background. While the simulated rate are obtained from the counts on small FOV detectors directly.

### 3.4 LE Backgrounds

The observed and simulated background spectra of LE are shown in Fig. 12. Only the small FOV detectors are used to obtain these spectra. Similar to ME, the observed spectra are from different periods after the launch, i.e. ∼ 6 months (20171103), 1 year (20180621) and 2 years (20190625). From the simulation, it can be seen that the CXB component dominates below ∼7.5 keV. While above ∼7.5 keV, most of the background results from cosmic ray protons. Contributions from the other components like SAA trapped protons and albedo gamma rays are negligible at the whole energy band. Therefore they are not presented in the plot.

The simulated background rate from 1 keV to 12 keV is ∼ $(12.8 \pm 0.2)$ counts s$^{-1}$ and the observed rate is $(12.1 \sim 12.7) \pm 0.1$ counts s$^{-1}$, which corresponds to a background level of $4.1 \sim 4.3 \times 10^{-3}$ counts s$^{-1}$ keV$^{-1}$ cm$^{-2}$. The long term variation is not seen due to the less contributions from radioactive components. The line features at 7.5 keV, 8.0 keV and 8.6 keV in the observed spectra are assumed to be fluorescence lines of nickel, copper and zinc, respectively. As mentioned in Section 2, zinc is the main composition of the LE aluminum alloy collimators. The 8.6 keV line of zinc is also presented in the simulated spectrum, but line structures at 7.4 keV and 8.0 keV are not evident. These discrepancies may result from the uncertainties of detector constructions in simulation. For instance, there is a glue layer between the SCD detectors and the ceramics, but the composition of this glue is unknown, therefore it is absent in the mass model. While it has high probability that this layer contains some elements that might show line structures in the background spectrum, as the glue does on ME background.

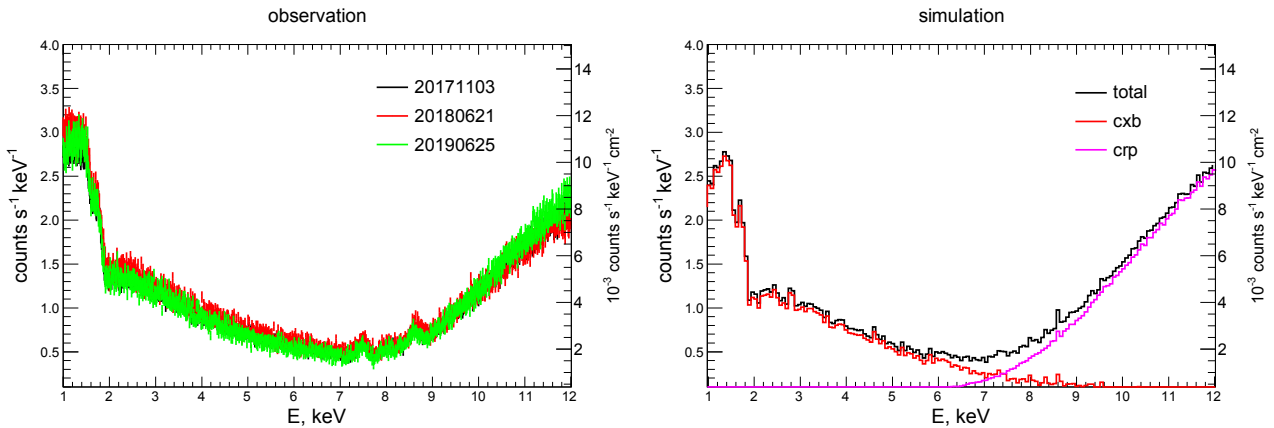

**Fig. 12** LE background spectra. Left: observed spectra. Lines are the same as in Fig. 11, which corresponds to an operation time of ∼ 6 months (20171103), 1 year (20180621) and 2 years (20190625) separately. Right: simulated spectra. For the simulation and observation spectra, only small FOV detectors are selected.

## 4 Concluding Remarks

The real-time background situation in space is much more complex than the simulation. Our simulation, with averaged space environment models and zenith satellite pointing attitude, could only give a general background estimation. On the one hand, for the space environments, the real time charged particles are not only modulated by solar activity and geomagnetic cutoff rigidity, but also affected by some short term turbulence. Take the LE background as an example, there are several hundred seconds flares below 7 keV presented in the light curve, which may result from the real-time low-energy charged particles (Liao et al. 2020). The observed ME background rates could vary by up to 50% even in the same COR region (Guo et al. 2020). In addition, the relative direction of the Earth in the FOV has some effects on the background. The Earth blocks CXB and primary cosmic rays, but it is also a radiator, especially for LE. That is why there is an additional selection condition that the bright earth angle is greater than 30°, as given in Table 2. On the other hand, the detector responses are also affected by the charge transfer and readout processes, especially for silicon detectors. There are already some available softwares that mimic these processes as realistically as possible, e.g the SIXTE software (Dauser et al. 2019). These processes are beyond the scope of Geant4, and we have not combined these processes currently.

In summary, we compare the simulation and the observation of *Insight-HXMT* backgrounds in this work. The simulations are based on Geant4 framework. The observation results are obtained from the analysis of the blank sky measurements. Deviations of the simulated background rates are within ∼ 20% from observations after the performance verification phase. For HE, this deviation approaches ∼ 5% after an operation time of two years, and the simulation well depicts the observed long-term increases of the background spectrum and the relative abundance of the ∼67 keV line. For ME and LE, long-term increases are not shown neither in simulation nor in observation, and the background levels given by simulations are also consistent with the observations. The predicted line structures in simulated background spectra, i.e. ∼31 keV, 56 keV, 67 keV and 191 keV lines of HE, the silver line of ME and the zinc line of LE, are observed after the launch. These lines are vital for the in-orbit calibration. For LE, the absence of fluorescence lines of nickel and copper in simulated spectra may result from the uncertainties of mass modeling. The agreement of these comparisons indicates that the space environment we adopted, physics processes selected and the detector constructions we built are reasonable in the *Insight-HXMT* mass model. The detailed differences between simulation and observation help us to study the low-Earth orbit space environment and the effect of detection process, which will be useful for the following EP and eXTP missions on instrument design, background rejection and estimation, etc.

**Acknowledgements** We thank the referee for helpful suggestions and comments. J. Zhang thanks Ying Tan for her help with the ME instrument performance and Helen Poon for improving the English of the whole text. This work is supported by the National Natural Science Foundation of China under the grant Nos. 11403026, U1838201, and U1938201. The observation data used in this work is from the *Insight-HXMT* mission, a project funded by the China National Space Administration (CNSA) and the Chinese Academy of Sciences (CAS). We gratefully acknowledge the support from the National Program on Key Research and Development Project (grant No. 2016YFA0400801) from the Minister of Science and Technology of China (MOST), and the Strategic Priority Research Program of the Chinese Academy of Sciences (grant No. XDB23040400).